# Effects of fear factors in disease propagation

Yubo Wang[1], Gaoxi Xiao[1]*, Limsoon Wong[2], Xiuju Fu[3], Stefan Ma[4] and Tee Hiang Cheng[1]

[1] School of Electrical and Electronic Engineering, Nanyang Technology University, Singapore 639798

[2] School of Computing and School of Medicine, National University of Singapore, Singapore 117543

[3] Institute of High Performance Computing, Singapore 138632

[4] Ministry of Health, Singapore 169854

Email: *egxxiao@ntu.edu.sg

**Abstract**. Upon an outbreak of a dangerous infectious disease, people generally tend to reduce their contacts with others in fear of getting infected. Such typical actions apparently help slow down the spreading of infection. Thanks to today's broad public media coverage, the fear factor may even contribute to prevent an outbreak from happening. We are motivated to study such effects by adopting a complex network approach. Firstly we evaluate the simple case where connections between individuals are randomly removed due to fear factor. Then we consider a different case where each individual keeps at least a few connections after contact reduction. Such a case is arguably more realistic since people may choose to keep a few social contacts, e.g., with their family members and closest friends, at any cost. Finally a study is conducted on the case where connection removals are carried out dynamically while the infection is spreading out. Analytical and simulation results show that the fear factor may not easily prevent an epidemic outbreak from happening in scale-free networks. However, it significantly reduces the fraction of the nodes ever getting infected during the outbreak.

## 1. Introduction

When a dangerous infectious disease is detected and known by public, especially revealed by the public media, people generally tend to reduce their contacts with others in fear of getting infected. Such reactions have been observed in past experiences, e.g., during the outbreak of Severe Acute Respiratory Syndrome (SARS) [1-2]. Intuitively it may be expected that such a fear factor should help reduce the number of infections or even prevent an epidemic outbreak from happening. Studies on such effects however are very limited. A related topic is to study on the spreading of rumors. While earlier work mainly focused on modeling rumor spreading itself [3-4], recent work evaluated the impacts of rumor spreading on disease control [5-6]. Funk et al. in [5] studied the case where positive information which lowers the susceptibilities of susceptible individuals spreads out in a compartmental model. It is found that in such a case, rumor helps reduce infection size yet does not lower the chance of epidemic outbreak. In [6], rumor helps immunize some susceptible individuals in well-mixed population. In such a model, not only the infection size is reduced, the chance of epidemic outbreak is also lowered if, and only if, rumor propagates fast enough.

We are motivated to evaluate the effects of fear factor. Our study is different from the existing ones in two aspects: first, we adopt a complex network approach, where individuals are modeled as nodes (vertexes) and the contacts between them for possible disease propagation are modeled as links (edges) [7-9]. It is well known that dynamics of epidemic spreading in networks can be strongly influenced by network topology [10-12]. As later we can see, adopting a complex network model leads to some different results from those in existing studies; second, we model the main effects of fear factor as leading to fewer connections between individuals rather than lowering or eliminating the susceptibilities of susceptible individuals. It may be easier to track the changes in social connections than to measure the changes in susceptibilities and arguably, reducing social connections may indeed be the main reason why fear factor helps lower infection size, especially in early stage of outbreak when cure or vaccination does not exist.

Studies on many real-life complex networks including human society and sexual contact networks show that they usually share some nontrivial common features. Among them the most noticeable one is probably that many of such networks are scale-free with their nodal degrees following a power-law distribution [13]. Specifically, the probability that a node is connected to $k$ other nodes is

$$P(k) \sim k^{-r}, \qquad (1.1)$$

where the exponent $r$ usually ranges between 2 and 3 [14-15].

In scale-free networks, a small number of high-degree nodes (hereafter termed as *hubs*) are connected to an extra large number of other nodes, making control of infectious disease highly difficult: once the hubs are infected, the infection can quickly spread out. Existing results show that epidemic spreading in an infinitely large scale-free network with an exponent $r \leq 3$ does not possess any positive *epidemic threshold*, defined as the phase-transition value of the infection's spreading capability below which the infection cannot cause a major epidemic outbreak [10-12, 16]. Having a zero epidemic threshold means that even a disease with weak transmissibility can easily survive and cause an outbreak. Further studies on the finite-size scale-free networks show that the epidemic threshold remains to be low and decreases slowly with an increasing network size [17].

In this paper, we study the dynamics of epidemic spreading in scale-free networks where some network links are removed due to the fear factor. The spreading of disease is assumed to follow the well-known Susceptible-Infected-Remove (SIR) scheme [18-19]. We evaluate three different cases as follows:

- Due to fear factor, a fraction of network links are randomly removed. We term such a case as with *random link removal*. It serves as a benchmark for other cases.
- The links are still randomly removed, however, subject to the constraint that after removals each node is still connected to at least a certain minimum number of links. Such is termed as *bounded random link removal*.
- Link removals from each node start only when infection spreading has reached the neighborhood of that node (to be defined in detail later); more infections in the neighborhood area trigger more link cuts. Such a model is termed as *dynamical link removal*, which may to some extent better resemble human behaviors when in face of a disease with relatively slow spreading.

Studies on the three simple yet typical cases provide some useful insights into the effects of the fear factor. Specifically, under random link removals, the inverse epidemic threshold is proportional to the portion of links being removed. To significantly increase the epidemic threshold, a large portion of links has to be removed. The link removals, however, help significantly reduce the number of nodes ever getting infected during the outbreak. In the bounded random link removal, subject to an overall number of links to be removed, keeping a minimum number of links for each node encourages lowering the number of links connected to hubs. Consequently, a higher bound value may actually lead to a higher epidemic threshold. In the dynamical link removal, the threshold is proportional to the percentage of link removal when there is one infection in the neighborhood. Fear factor in the three cases can significantly low infection sizes during outbreak. Note that the conclusions are different from those in [5]: though the nodes are not perfectly immunized, reducing social connections does help lower the epidemic threshold in any finite-size networks.

The remainder of this paper is organized as follows. In Section 2, we introduce the SIR model and the framework for analyzing SIR epidemic behaviors in contact networks. Section 3 studies the effects of random and bounded random link removals, respectively. In Section 4, dynamical link removal is defined in detail and its impacts on the epidemic spreading are evaluated. Finally, Section 5 concludes the paper.

## 2. SIR in contact networks

SIR model concisely describes the infectious disease propagation in human society through contacts between infective (I) individuals (those carrying the disease and capable of passing it to others) and susceptible (S) individuals (those who are healthy yet vulnerable to the disease). Infected individuals will eventually be removed (R) (either recovered and immunized or dead). The probability that an infectious individual eventually infects one of its neighbors is called the *transmissibility* of disease, denoted as $T$, which reflects the transmission capability of the disease [10]. There exists a critical value of transmissibility, known as epidemic threshold and denoted as $T_c$, below which the disease dies out exponentially rather than causing an outbreak. There are two different measures of infection size: *average outbreak size* defined as the average number of nodes ever getting infected before the disease dies out when $T < T_c$; and *epidemic size*, the average percentage of ever-infected nodes when $T \geq T_c$ [10].

A powerful tool for analyzing epidemic spreading in complex networks is *generating function* [10, 14-15], first introduced by Abraham de Moivre [20]. By encoding nodal-degree distribution and other relevant information into the coefficients of a power series, the generating function enables easier mathematical derivation. Specifically, in a random network with a degree distribution $P(k)$, following a randomly chosen link, the probability $q(k)$ of reaching a node connecting to $k$ other nodes is proportional to $(k+1)P(k+1)$ [10, 16]. $P(k)$ and $q(k)$ can be encoded into the probability generating function (PGF)

$$G_1(x) = \sum_{k=0}^{\infty} q(k) x^k = \frac{\sum_{k=0}^{\infty}(k+1)P(k+1)x^k}{\sum_{k=0}^{\infty}(k+1)P(k+1)} = \frac{G_0'(x)}{G_0'(1)}, \tag{2.1}$$

where

$$G_0(x) = \sum_{k=0}^{\infty} P(k) x^k. \tag{2.2}$$

The prime in the above equations denotes the derivative with respect to the argument. $G_0(x)$ and $G_1(x)$ hence contain the topological information of a random network with given nodal-degree distribution.

The SIR epidemic process on a network is equivalent to the bond percolation on the same network with a uniform bond occupation probability $T$ [10, 21-22]. Following the bond percolation theory, epidemic threshold can be calculated by [10]

$$T_c = \frac{1}{G_1^{'}(1)} = \frac{G_0^{'}(1)}{G_0^{''}(1)}. \tag{2.3}$$

For $T < T_c$, the average outbreak size is [10]

$$\langle s \rangle = 1 + \frac{TG_0^{'}(1)}{1 - TG_1^{'}(1)}. \tag{2.4}$$

And the epidemic size for $T \geq T_c$ is given as [10]

$$S = 1 - G_0(u), \quad \text{where} \quad u = 1 - T + TG_1(u). \tag{2.5}$$

The above existing results provide a framework for theoretical analysis, which will be adopted to analyze the epidemic threshold, the average outbreak size and the epidemic size in this paper.

## 3. Random and bounded random link removals

In this section, we study on the two different cases of random and bounded random link removals separately.

### *3.1. Random link removals*

Assume that the contact network is scale-free with nodal-degree distribution $P(k)$. Let a fraction $\rho$ of all the links be randomly removed from the network. Term the resulted network after link removals as the *subnet*. Apparently each link in the original network remains in the subnet with a probability $1-\rho$. Note that the subnet of a scale-free network after random link removal has the same asymptotic on degree distribution but does not exactly remain as a scale-free network [23-24].

Since a disease with transmissibility $T$ in the subnet can be modeled equivalently as a disease with transmissibility $T^* = (1-\rho)T$ in the original network, the epidemic threshold in the subnet can be expressed as

$$T_c^* = \frac{1}{1-\rho} T_c, \tag{3.1}$$

where $T_c$ is the epidemic threshold of the original network. Equation (3.1) reveals that the epidemic threshold of the subnet is inversely proportional to the portion of links removed. Random link removals therefore cannot easily increase the epidemic threshold: in an infinite scale-free network, the epidemic threshold remains as zero; in a finite scale-free network, even 50% link removals merely double the original epidemic threshold which is very low. Simulation results in a finite scale-free network are presented in figure 1. Throughout this paper, unless otherwise specified, numerical simulations are conducted on top of a 10,000-node scale-free network with degree distribution $P(k) = ck^{-3}$, the minimum nodal degree $k_{min} = 2$ and the maximum nodal degree $k_{max} = 100$. The network is generated by using the uncorrelated configuration model (UCM) [25].

The average outbreak size and epidemic size under transmissibility $T^*$ meanwhile can be calculated as

$$<s^*> = 1 + \frac{T(1-\rho)G_0'(1)}{1 - T(1-\rho)G_1'(1)} \quad (3.2)$$

and

$$S^* = 1 - G_0(u) \quad \text{where} \quad u = 1 - (1-\rho)T + (1-\rho)TG_1(u), \quad (3.3)$$

respectively. With a given transmissibility $T$, the epidemic threshold, average outbreak size and epidemic size in a random network with a given nodal degree distribution hence can be numerically solved as shown in figure 2.

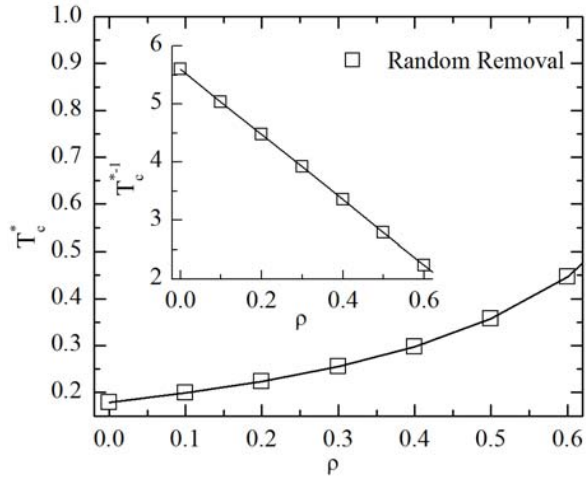

**Figure 1.** Dependence of epidemic threshold $T_c^*$ on the percentage of link removal $\rho$. The embedded sub-figure shows the relationship between the inverse epidemic threshold $T_c^{*-1}$ and $\rho$. The same symbols represent the corresponding inverse epidemic thresholds $T_c^{*-1}$ in the embedded sub-figures. Solid lines represent analytical results.

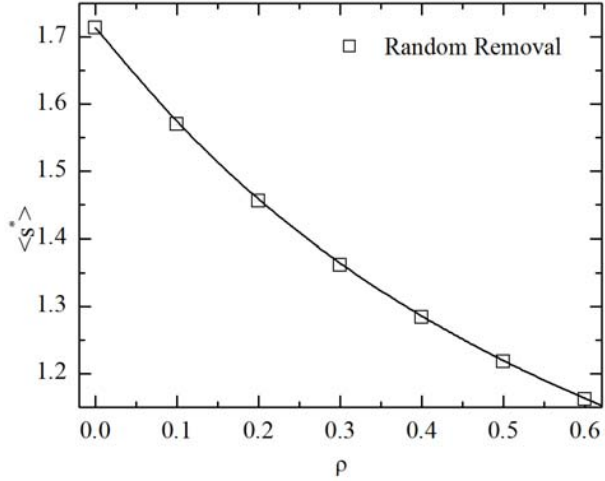

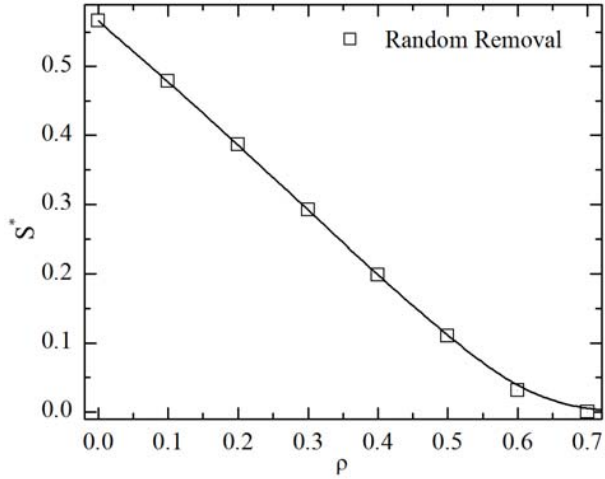

**Figure 2.** Dependence of infection size (average outbreak size $<s^*>$ in (a) and epidemic size $S^*$ in (b)) on the percentage of link removal $\rho$. Transmissibility is set at $T = 0.1$ in (a) and $T = 0.5$ in (b), respectively. Solid lines represent analytical results.

## *3.2. Bounded random link removal*

After random link removals, some network nodes may become isolated, which may not easily happen in real life, esp. in human societies. Instead, even when in panics, people typically still keep at least a few social links, e.g., with family members and close friends. To evaluate the effects of this factor, we study the bounded random link removal model as follows. For a network with $N<k>/2$ links, mark all these links as "uncolored" at the beginning. Then a total of $N<k>/2$ operations are carried out. In each operation, an "uncolored" link is randomly selected. It is removed at a probability of $\alpha$ if and only if both of its end nodes have degrees higher than a preset threshold value $k_B$. A selected but not removed link is marked as "colored". After $N<k>/2$ operations, each of the links in the original network is selected once and get either removed or colored. The resulted network composed of the colored links only is the subnet after bounded random link removals. Note that the random link removal can be viewed as a special case of the bounded random link removal where $k_B = 0$.

The link selection and removal operations can be viewed as being carried out in time steps. The degree distribution and the average degree of the contact network at time $t$ are denoted as $P(k,t)$ and $<k>_t$, respectively. For $t=0$, we have that $P(k,0) = P(k)$ and $<k>_0 = <k>$. At time $t$, the probability that a selected link is connected to a $(k+1)$-degree node is approximately $(k+1)P(k+1,t)/<k>_t$. Taking down such a link decreases the number of nodes with degree $k+1$ while increasing the number of degree-$k$ nodes. By treating $P(k,t)$ as a function of $t$, we have that at each step a link is removed at a probability of

$$P_r(t) = \left(1 - \frac{1}{<k>_t} \sum_{k=0}^{k \leq k_B} kP(k,t)\right)^2 \cdot \alpha. \tag{3.4}$$

The overall portion of link removal, still denoted as $\rho$, can be derived as

$$\rho = \frac{2}{N<k>_0} \sum_{t=1}^{N<k>_0/2} P_r(t) \tag{3.5}$$

In each time step, the degree distribution of the contact network evolves approximately as

$$P(k,t+1) = P(k,t) \qquad \text{for } k < k_B, \tag{3.6}$$

$$P(k,t+1) = P(k,t) + \frac{2P_r(t)(k+1)P(k+1,t)}{N(<k>_t - \sum_0^{k_B} kP(k,t))} \qquad \text{for } k = k_B, \tag{3.7}$$

$$P(k,t+1) = P(k,t) + \frac{2P_r(t)((k+1)P(k+1,t) - kP(k,t))}{N(<k>_t - \sum_0^{k_B} kP(k,t))} \qquad \text{for } k > k_B. \tag{3.8}$$

And the average nodal degree evolves as $<k>_{t+1} = <k>_t - 2P_r(t)/N$. The final degree distribution of the contact network for bounded random link removal can be calculated by iterating equations (3.6)-(3.8) for $N<k>_0/2$ times. The epidemic threshold and infection sizes then can be calculated using equations (2.3)-(2.5).

Figure 3 shows the analytical and simulated degree distributions of contact network before and after bounded random link removals respectively. The densities of nodes at lowest and highest degrees change significantly, leading to different epidemic behaviors of the networks.

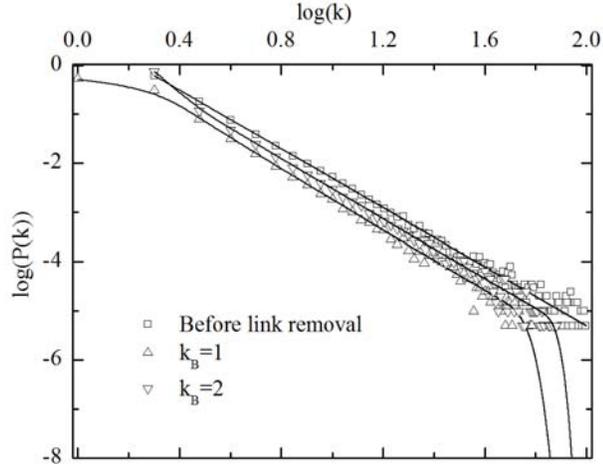

**Figure 3.** Degree distributions of a scale-free network before and after bounded random link removals respectively. $\alpha$ is set to be 0.5. The square boxes represent the degree distribution before link removal; up and down triangles represent degree distributions after bounded random link removal with bounds being 1 and 2, respectively. The solid lines show the corresponding analytical results.

It is worth noting that two approximations are adopted in equations (3.6)-(3.8), which help keep equations and calculations rather simple yet may affect the accuracy of the analysis, especially when $\alpha$ is of a large value (e.g., close to 1). The first approximation is to assume that the network remains to be random after bounded random link removals. In fact, since links connecting any nodes with degrees lower than $k_B$ are not removed, the randomness of the network does not strictly hold. Figure 4 shows the changes of network assortativity coefficient (as defined in [26]) during the process of bounded random link removals under different values of $\alpha$ and $k_B$. As we know, the assortativity coefficient should be zero in a random network, which is not the case in figure 4 when $\alpha = 1$. The degree-degree correlation in the subnet however only becomes significant when $\alpha$ is quite large (close to 1). For a moderate value where $\alpha = 0.5$, the assortativity coefficient remains to be close to zero.

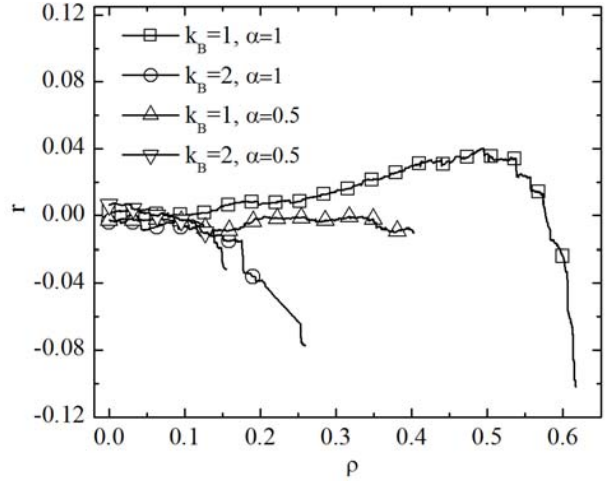

**Figure 4.** Network assortativity coefficient $r$ under different values of $\alpha$ in the process of bounded random link removals. Square boxes and circles represent the cases where $k_B = 1$ and $k_B = 2$ respectively while $\alpha = 1$. Up triangles and down triangles indicate the simulation results for corresponding cases while $\alpha = 0.5$.

Effects of the second approximation can be observed in figure 5, which reveals the relationship between the value of $\alpha$ and the overall percentage of link removal $\rho$. We observe that there is a good match between analytical and simulation results when $\alpha$ is of a small or moderate ($\alpha \leq 0.5$) value. When $\alpha$ gets close to 1, however, the simulation results of $\rho$ become higher than the analytical results. This can be explained. In simulation, a colored but not removed link will not be selected ever again, while in analysis, to keep the equations reasonably simple, a colored link, as long as it is not removed, may be selected again. For small values of $\alpha$, differences between analytical and simulation results are trivial. When $\alpha$ gets larger, however, the differences become no longer negligible.

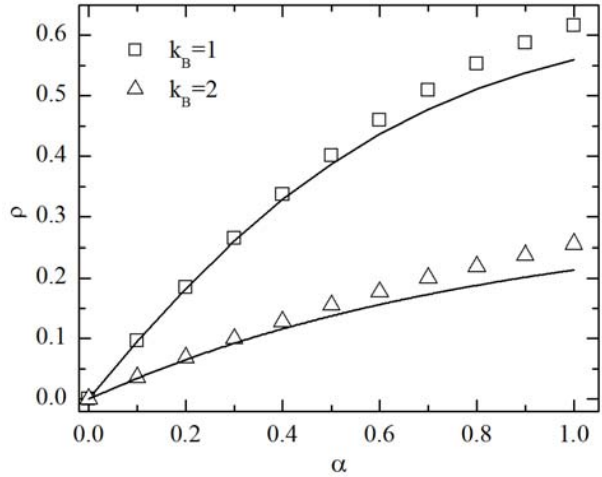

**Figure 5.** Relationship between the value of $\alpha$ and the overall percentage of link removal $\rho$ in the bounded random link removal. The square boxes and the up triangles represent simulation results where the lower bounds are set as 1 and 2, respectively. The solid lines present the analytical results.

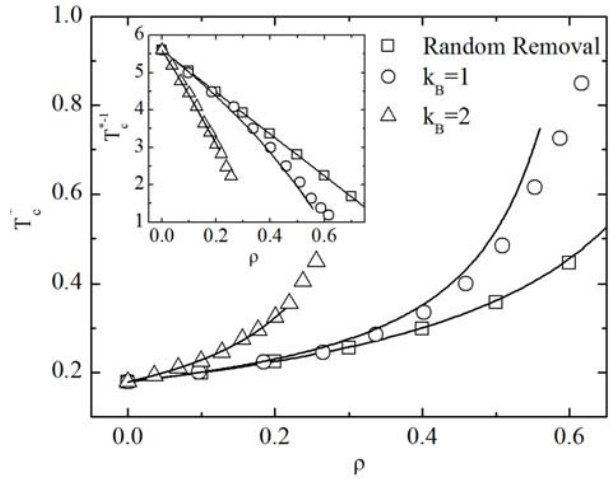

Figure 6. Dependence of epidemic threshold $T_c^*$ on the percentage of link removal $\rho$. The embedded sub-figure shows the relationship between the inverse epidemic threshold $T_c^{*-1}$ and $\rho$. The square boxes represent the results for the case of random link removal. Circles and up triangles show the results under bounded random link removal with different lower bounds $k_B$. Solid lines present the corresponding analytical results.

Figure 6 reveals the relationship between the epidemic threshold and link removal percentage $\rho$. We see that subject to a given $\rho$, bounded random link removals lead to a higher epidemic threshold than that of the case under random link removals, and a higher value of $k_B$ leads to a higher epidemic threshold. This can be explained: subject to a given percentage of overall link removal, having a higher value of $k_B$ encourages more links connected to high-degree nodes to be removed. As a result, the epidemic threshold is increased. Such an observation may have significance in reality: keeping a few links with family members and close friends may make losing connections with socially active friends more bearable and consequently helps control disease spreading.

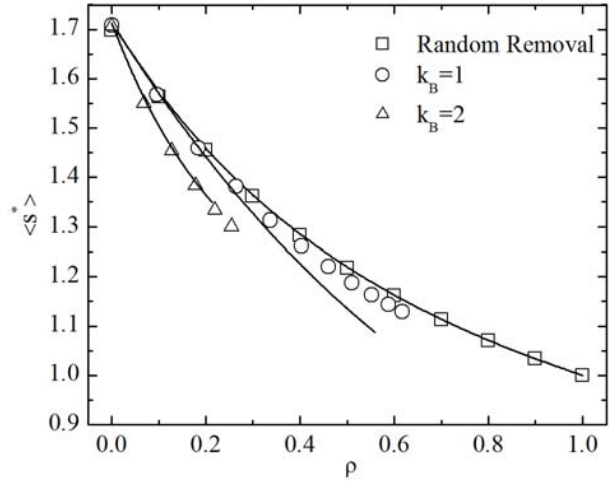

**Figure 7.** Dependence of the average outbreak size $<s^*>$ on the percentage of link removal $\rho$. The epidemic transmissibility is set at $T = 0.1$. The square boxes are for the case of random link removal. Circles and up triangles show the results for the case after the bounded random link removal with different lower bounds $k_B$. Solid lines represent the analytical results.

Figures 7 and 8 show analytical and simulation results of infection sizes under disease spreading with different transmissibility. We see that in all the different cases, link removals help significantly reduce the infection sizes. As to the effects of the lower bound $k_B$, a higher lower-bound value leads to a smaller infection size when the disease is not extremely infectious, as we have discussed earlier. When the transmissibility is very strong, however, as shown in figure 8(c), a higher $k_B$ may lead to a slightly larger infection size. This can be understood: to keep safe from a strongly infectious disease, most or all connections have to be removed. Even keeping only a few connections may put the node at a high risk of getting infected. Note that a high value of $k_B$ restricts the maximum link removal probability that can be achieved, which may weaken network protection against strongly infectious diseases. For example, in the simulated network, when $k_B=2$, $\rho$ cannot be larger than 0.214. Also note that larger differences between analytical and simulation results may be observed in figures 7 and 8 when $\rho$ is of larger values, of which the reasons have been discussed earlier.

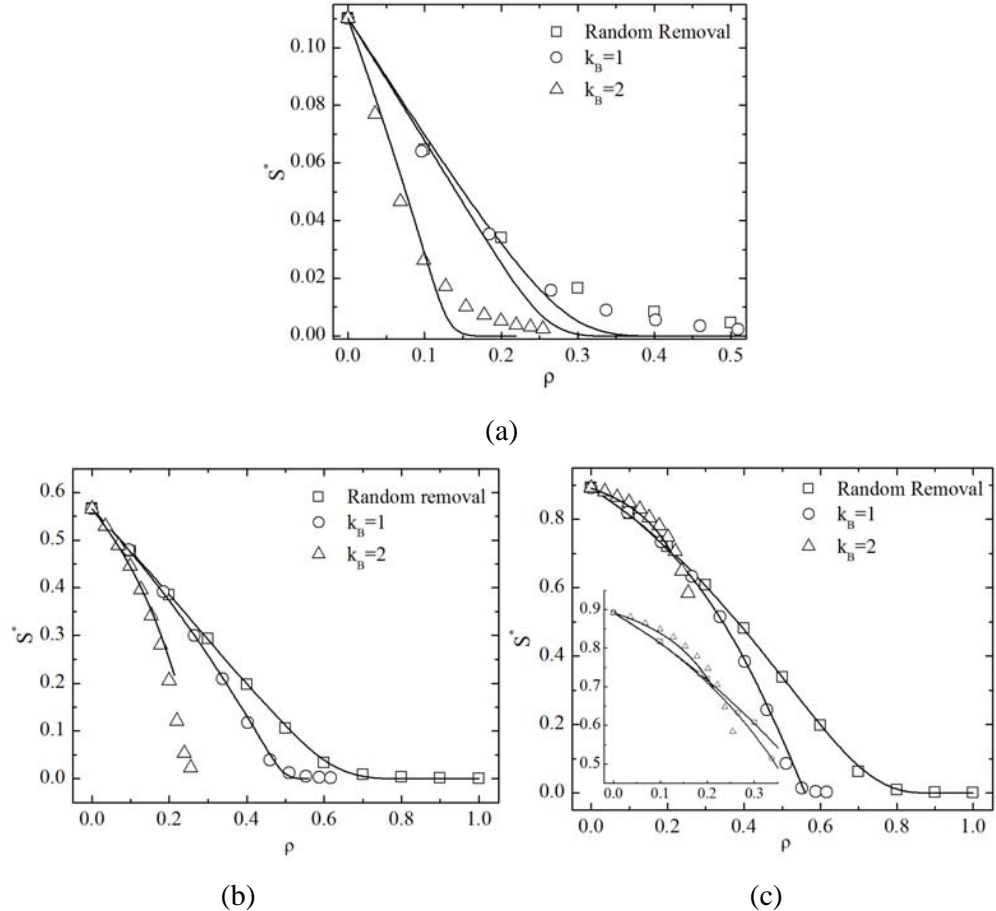

(a)

(b)                                          (c)

**Figure 8.** Dependence of epidemic size $S^*$ on the percentage of link removal $\rho$. The epidemic transmissibility is set at $T=0.25$, $T=0.5$ and $T=0.75$ in sub–figures (a), (b) and (c), respectively. The square boxes are for the case of random link removal; while circles and up triangles represent simulation results for the cases after the bounded random link removal with different lower bounds $k_B$. Solid lines represent corresponding analytical results.

## 4. Dynamical link removal

In Section 3, it was assumed that the network topology remains static after link removals. In reality, however, people may tend to cut more social connections when there are more infected cases around. In this section, we evaluate the effects of such dynamic link removals.

We study on the model as follows: the disease spreads in a network following the SIR model with slotted time. In each time slot, an infected node transmits disease to its susceptible neighbor nodes with probability $T$; infected node will become removed within unity time [27]. Due to the fear factor, each susceptible node removes a portion of $f(n) = 1 - A\beta^n$ of its links, where $n$ is the number of infected and removed nodes around, $A$ and $\beta$ are parameters between 0 and 1. Smaller values of $A$ and $\beta$ lead to a larger portion of link removal. Note that the special case where $\beta = 1$ is equivalent to the random link removal discussed in Section 3.1, where links are removed at a portion of $1 - A$.

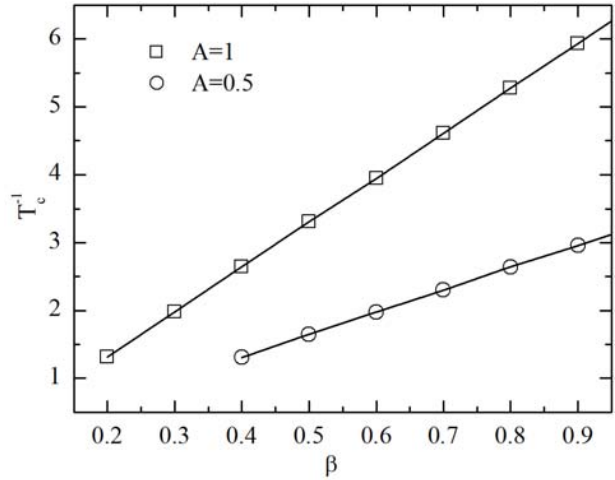

**Figure 9.** Linear relationship between the inverse epidemic threshold and the parameters $A$ and $\beta$. Square boxes represent the results where $A = 1$; circles are for the case where $A = 0.5$. The solid straight lines are calculated using the least square estimate.

Let $T_c$ be the epidemic threshold of the original network without link removal. To analyze epidemic threshold under the dynamic link removal, we consider the case with a low transmissibility $T$ close to the epidemic threshold in a large network. Since the number of infections under such case is very low, the probability of having two or more infected nodes connected to the same susceptible node can be neglected. Each susceptible node can be regarded as connected to at most one infected node, which causes a fraction of $1 - A\beta$ of its links to be removed and consequently reduces the chance of getting infected to $A\beta T$. Therefore the epidemic threshold $T_d$ is increased to

$$T_d = T_c / (A\beta), \tag{4.1}$$

which is inversely proportional to $A\beta$. Such a relationship can be observed in figure 9.

The topology of the contact network keeps changing during dynamical link removals. To reveal the infection size along time, we study the model at the mean-field level [28] (A more sophisticated yet more precise approach is to calculate the exact topology of contact network in each time step and then analyze the distribution of infection based on it. Such an approach has been adopted in [27]. In this paper, we focus on evaluating the effects of fear factors rather than proposing most accurate analysis. Thus we propose simple but still reasonably accurate analysis.). Denote the densities of susceptible, infected and removed $k$-degree nodes at time $t$ as $I_k(t)$, $S_k(t)$ and $R_k(t)$, and their respective initial conditions as $I_k(0)$, $S_k(0)$ and $R_k(0)$. It can be derived that

$$I_k(t+1) = S_k(t) \sum_{k'=0}^{k} \binom{k}{k'} \theta_{R+I}^{k'}(t)(1-\theta_{R+I}(t))^{k-k'}$$

$$\sum_{k''} \binom{k'}{k''} (\theta_I(t)/\theta_{R+I}(t))^{k''} (1-\theta_I(t)/\theta_{R+I}(t))^{k'-k''} (1-(1-T(1-f(k')))^{k''}), \quad (4.2)$$

$$S_k(t+1) = S_k(t) - I_k(t+1),$$

$$R_k(t+1) = R_k(t) + I_k(t),$$

where $\theta_I(t)$ denotes the probability that a randomly selected link is connected to an infected node at time $t$, and $\theta_{R+I}(t)$ the probability that a randomly selected link connects to an infected or a removed node. We have

$$\theta_I(t) = \frac{1}{<k>} \sum_k k P(k) I_k(t), \quad (4.3)$$

$$\theta_{R+I}(t) = \frac{1}{<k>} \sum_k k P(k) (I_k(t) + R_k(t)). \quad (4.4)$$

In equation (4.2), the term $\binom{k}{k'} \theta_{R+I}^{k'}(t)(1-\theta_{R+I}(t))^{k-k'}$ calculates the probability that a $k$-degree node is connected to a total of $k'$ infected or removed nodes; while $\binom{k'}{k''}(\theta_I(t)/\theta_{R+I}(t))^{k''}(1-\theta_I(t)/\theta_{R+I}(t))^{k'-k''}$ shows the probability that among these $k'$ nodes, there are $k''$ infected nodes. If a susceptible node is connected to a total of $k'$ infected and removed nodes among which $k''$ nodes are infected ($k'' \leq k'$), it will be infected at a rate of $1-(1-T(1-f(k')))^{k''}$. Equation (4.2) calculates how the $k$-degree nodes change between different statuses in each time slot. The density of the infected nodes in the network at time $t$ therefore can be expressed as $I(t) = \sum_k P(k) I_k(t)$, while the densities of susceptible and removed nodes can be calculated similarly.

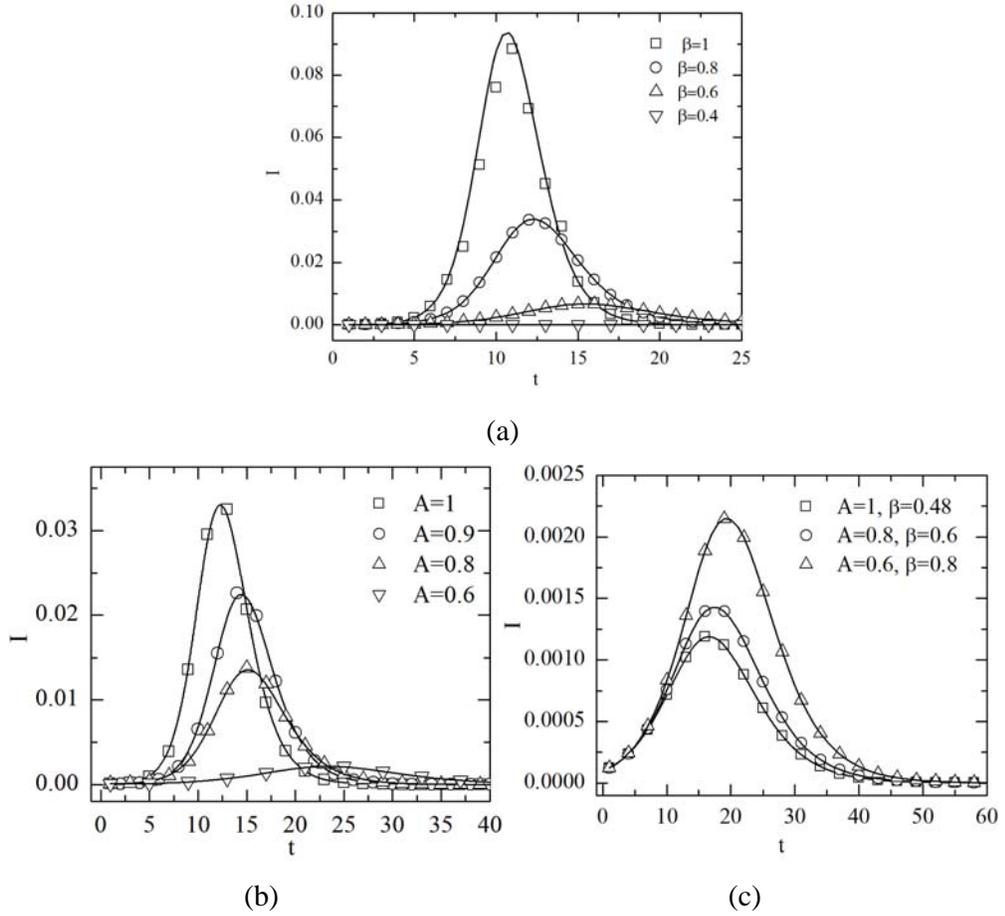

**Figure 10.** Densities of infected nodes along time under dynamic link removals. In (a), the disease transmissibility $T=0.4$ and $A=1$. In (b), $T=0.4$ and $\beta=0.8$. In (c), $T=0.4$. The four symbols show simulation results under different values of $\beta$ or $A$. The solid lines show the analytical results.

Figure 10(a) illustrates the sizes of infected nodes along time with different values of link removal parameter $\beta$. $\beta=1$ corresponds to the case of random link removal, leading to the largest infection size and earliest arrival of the peak of infections (i.e., the time with the largest infection size). A smaller value of $\beta$ leads to a significantly smaller infection size, as well as a longer time before the infection size reaches its peak. Such observations show that, if fear gets stronger when there are more infected cases in the neighborhood, leading to more link cuts, the infection size can be significantly reduced and more time may be bought before the worst infection peak arrives. With a given value of $\beta$, a smaller value of $A$, which corresponds to a stronger fear effect, also leads to a much smaller infection size and a long time before the infection peak arrives, as we can easily observe in figure 10(b). Finally, figure 10(c) shows the results for the case where $A\beta$ remains as a constant. From equation (4.1) we know that under such case the epidemic threshold remains as a constant. However, the infection sizes corresponding to different combinations of $A$ and $\beta$ may be very different. As we can see, generally a smaller value of $\beta$ leads to a smaller infection size yet slightly earlier arrival of the infection peak, This can be easily understood: subject to a given value of $A\beta$, a smaller value of $\beta$ lets more links connected to a susceptible node to be removed, which makes the spreading of infection harder and reach its peak earlier. Note that when transmissibility $T$ is very low close to $T_c$, different combinations of $A$ and $\beta$ make only marginal differences since the probability that a susceptible node is connected to two or more infected or removed node can be neglected under such case [26].

## 5. Summary and Discussion

Random, bounded random and dynamical link removal models were proposed for evaluating the effects of reducing contacts in fear of getting infected during an epidemic outbreak. We found that under random link removal, there exists a linear relationship between the inverse epidemic threshold and the portion of link removed, which suggests that fear factor alone cannot easily prevent an outbreak from happening in scale-free networks. However, the infection size can be significantly reduced even when only a small portion of links are taken down.

Bounded random link removal restricts the maximum percentage of links that can be removed. Subject to a link removal percentage that can be reached, having a higher value of the lower bound encourages removing more links with high-degree nodes and consequently helps restrict disease spreading. When the disease is extremely infectious, however, bounded random link removal is not as effective as random link removal.

The dynamic link removal model has an epidemic threshold inversely proportional to the product of link removal parameters $A$ and $\beta$. A smaller value of $A$ and $\beta$ generally leads to a significantly smaller infection size, and buys more time before the arrival of the peak of infection as well. Subject to a given product of $A$ and $\beta$, a smaller value of $\beta$ leads to a much smaller infection size and slightly earlier arrival of the peak of infection.

Many interesting further studies can be carried out on the effects of fear factors. For example, in the bounded random link removal model, it was assumed that low-degree nodes can veto the removals of links, including those with high-degree nodes. In real life, this may not be the case. They may not have veto power. Also, people may be very willing to remove their links with highly connected individuals since the chance that such individuals are infected is high. To compensate the loss of social connections, people may actually strengthen or even build up new connections with less socially active ("safer") individuals. While connections with high-degree nodes are quickly removed, connections between low-degree nodes may in some cases be enhanced. It is not clear yet the effects of such changes.

In this paper, contact networks are assumed to be random networks. In real life, however, most networks are not strictly random. They may show either assortative or disassortative mixing properties [26], and with various types of community structures. Inter-community and intra-community links could have different impacts on the disease propagation [29], and such links may be removed in different ways due to fear factor. All these will be of our future research interest.

## Acknowledgement

This work is supported in part by Ministry of Education, Singapore, under grant RG27/09.